\documentclass[twocolumn,aps,floatfix,pra,showpacs]{revtex4}
\usepackage{amssymb,amsfonts,mathrsfs}
\usepackage{graphicx,bm}
\usepackage[T1]{fontenc}
\begin{document}

\title{Spinor condensate of $^{87}$Rb as a dipolar gas}
\author{Tomasz \'Swis\l ocki$\,^{1}$, Miros\l aw Brewczyk$\,^2$, Mariusz Gajda$\,^{1}$, and Kazimierz Rz\k{a}\.zewski$\,^{3,}\,^{4}$}
\affiliation{
\mbox{$^1$ Instytut Fizyki PAN, Al.Lotnik\'ow 32/46, 02-668 Warsaw, Poland}
\mbox{$^2$Wydzia\l {} Fizyki, Uniwersytet w Bia\l ymstoku, ul. Lipowa 41, 15-424 Bia\l ystok, Poland}
\mbox{$^3$ Wydzia\l {} Matematyczno-Przyrodniczy SN\'S UKSW, ul. Dewajtis 5, 01-815 Warsaw, Poland}
\mbox{$^4$ Centrum Fizyki Teoretycznej PAN, Al.Lotnik\'ow 32/46, 02-668 Warsaw, Poland}}
\date{\today}
\begin{abstract}
We consider a spinor condensate of $^{87}$Rb atoms in $F=1$ hyperfine state confined in an optical dipole trap. Putting initially all atoms in $m_F=0$ component we find that the system evolves towards a state of thermal equilibrium with kinetic energy equally distributed among all magnetic components. We show that this process is dominated by the dipolar interaction of magnetic spins rather than spin mixing contact potential. Our results show that because of a dynamical separation of magnetic components the spin mixing dynamics in $^{87}$Rb condensate is governed by dipolar interaction which plays no role in a single component rubidium system in a magnetic trap.
\end{abstract}
\maketitle
Since the first achievement of a Bose-Einstein condensation of trapped atomic gases, an ultracold atomic $^{87}$Rb vapor is considered as the generic example of an atomic condensate. Rubidium atoms have many features which make them a perfect candidate for ultracold atoms experiments. At temperatures in the nanokelvin regime, interatomic interactions can be, to a very high accuracy, approximated by a short range two body potential with the s-wave scattering length $a_s$ being the only relevant parameter. In a magnetic trap, the two low field seeking states stable with respect to two body processes, are the two hyperfine states of the atomic rubidium ground state: $|F=2, m_F=2\rangle$ or $|F=1, m_F=-1\rangle$. Simple estimation of a characteristic contact energy gives:
\begin{equation}
\label{contact}
E_c=\left( 4 \pi \hbar^2 a_s/M\right) n\,,
\end{equation}
where $a_s$ is the s-wave scattering length, $M$ is the mass of the atom and $n$ is a typical atomic density. The spin dipole-dipole interaction energy is of the order of:
\begin{equation}
\label{dipole}
E_d=\mu^2 n\,,
\end{equation}
where magnetic moment of $^{87}$Rb in an $F=1$ hyperfine state is equal to $\mathbf{\mu}=\frac{1}{2}\mu_B$ where $\mu_B$ is the Bohr magneton. The ratio of these two energies is:
\begin{equation}
E_d/E_c = 4.2\times10^{-4}\,.
\end{equation}
Due to the smallness of the atomic magnetic moment the dipole-dipole interaction between atoms is several orders of magnitude smaller than the contact interaction and can be ignored. The rubidium condensate in a magnetic trap is thus a perfect example of a weakly interacting system with contact interaction characterized by a single parameter -- the scattering length.

Experimental achievement of a condensate of atoms with sizeable long range dipolar forces was a great challenge. Finally, after overcoming a number of serious difficulties a chromium $^{52}$Cr condensate characterized by a relatively large dipole moment was created \cite{Griesmaier}. Moreover, by utilizing the technique of Feshbach resonance the contact term was practically turned off and pure dipolar condensate was achieved \cite{Pfau_Koch}.

In this paper we show that yet another and experimentally much simpler way to dipolar condensates is possible. Namely, optical dipole traps allow for a simultaneous trapping of various magnetic components of a given hyperfine state, i.e. spinor condensates. A spin dynamics of rubidium $F=1$ and $F=2$ states and the formation of a condensate due to increase of the atomic number in a given Zeeman sublevel was studied experimentally \cite{Sengstock_1}. In the experiment \cite{Chapman} a transfer of atoms from the initial $m_F=0$ state to $m_F=\pm 1$ states was observed.

The theoretical studies related to \cite{Chapman} were performed in one or two spatial dimensions \cite{Lewenst,Saito,Gawryluk_1} and dipole-dipole interactions so far were ignored. On the other hand, some theoretical and experimental evidence of enhanced role of dipolar forces in $F=1$ $^{87}$Rb was reported. This is the observation of disintegration of a helical structure of magnetization \cite{Kurn_prl} or Einstein-de Haas effect \cite{Ueda_3,Gajda_Bongs}. Some authors \cite{Ueda_1,Kurn_prl,Pu} have already stressed an enhanced role of dipole-dipole interactions in a spin dynamics of ferromagnetic $^{87}$Rb. This is because the contact spin mixing term (see the Hamiltonian Eq. (\ref{hamiltonian})) is proportional to $E_{s}=4 \pi \hbar^2 (a_2-a_0)n/3M$ where the scattering lengths $a_0=5.387$nm and $a_2=5.313$nm determine collisional cross sections in a channel of a total spin 0 and 2 respectively. Therefore a ratio of the dipolar energy to the spin mixing contact term in the ferromagnetic rubidium is as large as: 
\begin{equation}
E_{d}/E_{s}=0.09\,.
\label{e_spin}
\end{equation}
In order to analyze processes responsible for a spin dynamics we shall discuss a role of different terms of the Hamiltonian of the system:
\begin{eqnarray}
\hat{H} &=& \int d^3r \left( \hat{\Psi}_i^{\dagger}(\mathbf{r}) H_0^{} \hat{\Psi}_i^{}(\mathbf{r}) 
- \gamma \hat{\Psi}_i^{\dagger}(\mathbf{r}) \mathbf{B} \mathbf{F}_{ij} \hat{\Psi}_j^{}(\mathbf{r}) \right. \nonumber \\
&+& \left. \frac{1}{2} c_0^{} \hat{\Psi}_j^{\dagger}(\mathbf{r}) \hat{\Psi}_i^{\dagger}(\mathbf{r}) \hat{\Psi}_i^{}(\mathbf{r}) \hat{\Psi}_j^{}(\mathbf{r}) \right. \nonumber \\
&+& \left. \frac{1}{2} c_2^{} \hat{\Psi}_k^{\dagger}(\mathbf{r}) \hat{\Psi}_i^{\dagger}(\mathbf{r}) {\bf F}_{ij} {\bf F}_{kl} \hat{\Psi}_j^{}(\mathbf{r}) \hat{\Psi}_l^{}(\mathbf{r}) \right) \nonumber \\
&+& \frac{1}{2}\int d^3r  d^3r' \hat{\Psi}_k^{\dagger}(\mathbf{r}) \hat{\Psi}_i^{\dagger}(\mathbf{r'}) V_{ij,kl}^{d} (\mathbf{r}-\mathbf{r'}) \hat{\Psi}_j^{}(\mathbf{r'}) \hat{\Psi}_l^{}(\mathbf{r})\, \nonumber \\
&&
\label{hamiltonian}
\end{eqnarray}
where repeated indices (taking values +1,0 and -1) are to be summed over. 

The first term in (\ref{hamiltonian}) is the single particle kinetic energy and the trapping potential energy $V_{tr}$: $H_0=-\frac{\hbar^2}{2 M} \nabla^2 + V_{tr}(\mathbf{r})$. The second term describes the interaction with magnetic field $\mathbf{B}$ with $\gamma$ being the gyromagnetic coefficient which relates the effective magnetic moment with the hyperfine angular momentum $(\boldsymbol{\mu}=\gamma \mathbf{F})$. The terms with coefficients $c_0 = 4 \pi \hbar^2 (a_0+2 a_2)/3M $ and $c_2 = 4 \pi \hbar^2 (a_2-a_0)/3M $ describe the spin-independent and spin-dependent parts of the contact interactions. Because $c_2 < 0$ the ground state of ${}^{87}$Rb is ferromagnetic. $\mathbf{F}$ are the spin-1 matrices. The last term describes the magnetic dipolar interaction of two magnetic dipole moments located at $\mathbf{r}$ and $\mathbf{r'}$: 
\begin{eqnarray}
V_{ij,kl}^{d} (\mathbf{r}-\mathbf{r'}) &=& \frac{\gamma^2}{\left|\mathbf{r}-\mathbf{r'}\right|^3} \mathbf{F}_{ij} \mathbf{F}_{kl} - \frac{3 \gamma^2}{\left|\mathbf{r}-\mathbf{r'}\right|^5} \times \nonumber \\ && \left( \mathbf{F}_{ij} \cdot (\mathbf{r}-\mathbf{r'}) \right) \left(\mathbf{F}_{kl} \cdot (\mathbf{r}-\mathbf{r'}) \right)\,.
\label{dip}
\end{eqnarray}

The field operator $\hat{\Psi}_i(\mathbf{r})$ annihilates an atom in the hyperfine state $|F=1,i\rangle$ at a point $\mathbf{r}$. Using the classical fields approximation \cite{r1} we replace the field operators $\hat{\Psi}_{i}^{}(\mathbf{r})$ by the classical wavefunctions $\Psi_{i}^{}(\mathbf{r})$. The equation of motion for these wavefunctions is
\begin{eqnarray}
i \hbar \frac{\partial}{\partial t} \left(\begin{array}{c}
{\Psi}_{1}^{} \\
{\Psi}_{0}^{} \\
\hspace{0.25cm} {\Psi}_{-1}^{}
\end{array} \right) = \left(H_{0} + H_{B} + H_{c} + H_{d} \right) \left(\begin{array}{c}
{\Psi}_{1}^{} \\
{\Psi}_{0}^{} \\
\hspace{0.25cm} {\Psi}_{-1}^{}
\end{array} \right).
\label{ruch}
\end{eqnarray}

The diagonal part of $H_c$ is given by $H_{c11}=(c_0+c_2) {\Psi}_{1}^{*} {\Psi}_{1}^{} + (c_0+c_2){\Psi}_{0}^{*} {\Psi}_{0}^{} + (c_0-c_2){\Psi}_{-1}^{*} {\Psi}_{-1}^{}$,  $H_{c00}=(c_0+c_2) {\Psi}_{1}^{*} {\Psi}_{1}^{} + c_0{\Psi}_{0}^{*} {\Psi}_{0}^{} + (c_0+c_2){\Psi}_{-1}^{*} {\Psi}_{-1}^{}$, $H_{c-1-1}=(c_0-c_2) {\Psi}_{1}^{*} {\Psi}_{1}^{} + (c_0+c_2){\Psi}_{0}^{*} {\Psi}_{0}^{} + (c_0+c_2){\Psi}_{-1}^{*} {\Psi}_{-1}^{}$. The off-diagonal elements that describe collisions not preserving the spin projection of each atom are equal to $H_{c10} = c_2 {\Psi}_{-1}^{*}{\Psi}_{0}^{}$, $H_{c0-1} = c_2 {\Psi}_{0}^{*}{\Psi}_{1}^{}$. Moreover $H_{c1-1} =0$. On the other hand, for the $H_d$ term one has $H_{dij} = \int d^3r' {\Psi}_{n}^{*} (\mathbf{r'}) V_{ij,nk}^{d} {\Psi}_{k}^{} (\mathbf{r'})$. This term is responsible for the change of total spin projection of colliding atoms. 

In the following we use the oscillator units where a distance is measured in $a_{h0}=\left(\hbar/M\omega\right)^{1/2}$ where $\omega=2\pi \times 100$ Hz. Initially the system is excited with all atoms in $m_F=0$ component (with $m_F=\pm 1$ components equal to zero). As the first step we compute the $m_F=0$ ground state of the system by means of the imaginary time propagation. The resulting wave function was then randomly perturbed in order to inject about $10\%$ of the excess energy. 

The details of the classical field approximation are reviewed in \cite{r1}. 
\begin{figure}[!htb]
\begin{center}
\resizebox{3.5in}{2.0in} {\includegraphics{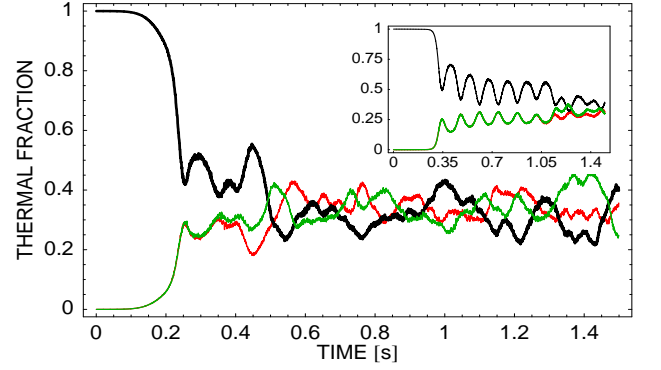}}
\caption[]{(color online) Populations of thermal clouds for $m_F=+1$ (red), $ m_F=0$ (black), and $ m_F=-1$ (green) states as a function of time with (the main frame) and without (the inset) dipole-dipole interactions. The parameters are N=$3$x$10^5$, $\beta=1$. Total populations of $m_F=\pm 1$ components are identical if dipole-dipole interaction is turned off. The external magnetic field is equal to zero.
\label{nor_sf_300k}}
\end{center}
\end{figure}
\begin{figure}[!htb]
\begin{center}
\resizebox{3.2in}{1.85in} {\includegraphics{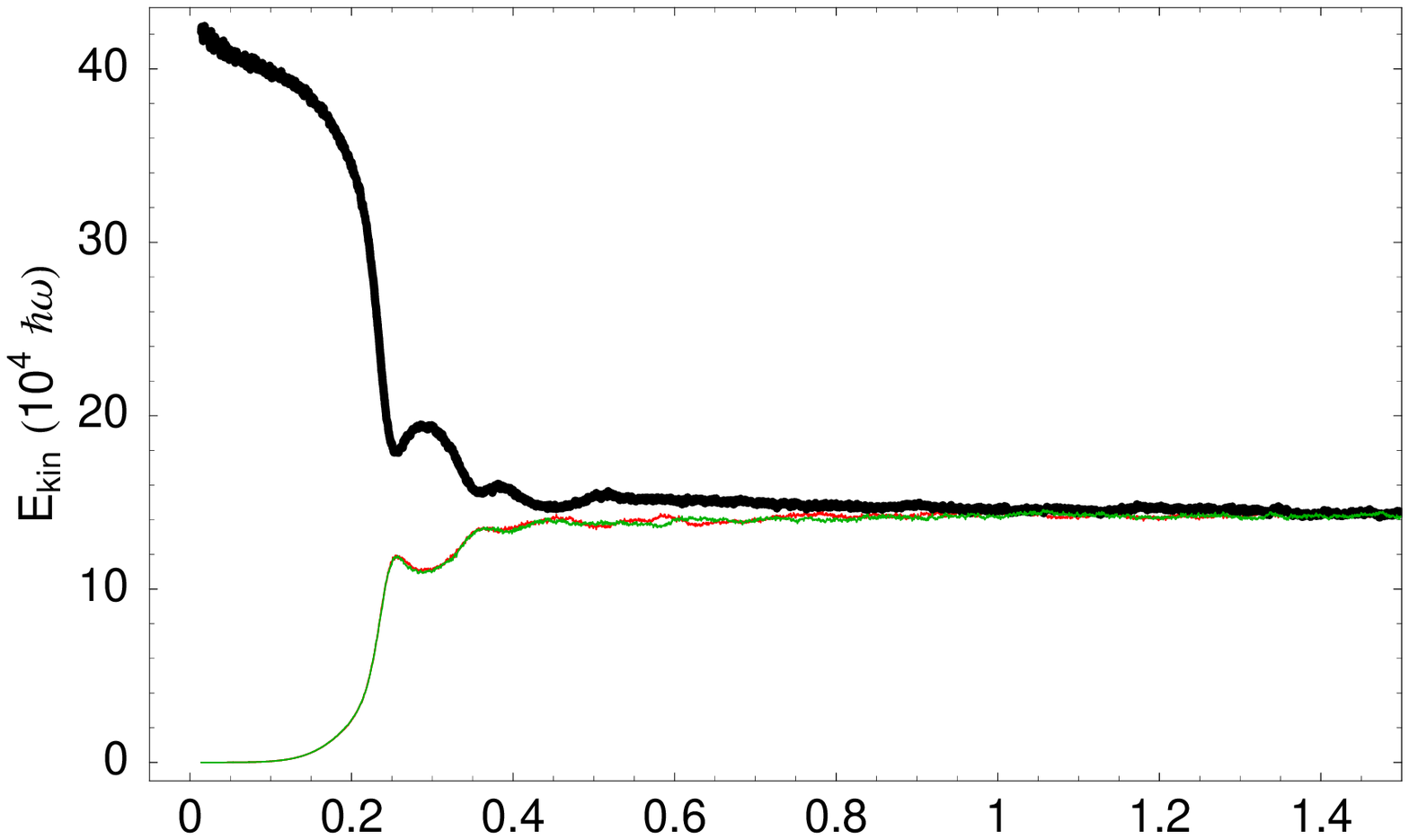}}
\resizebox{3.2in}{1.85in} {\includegraphics{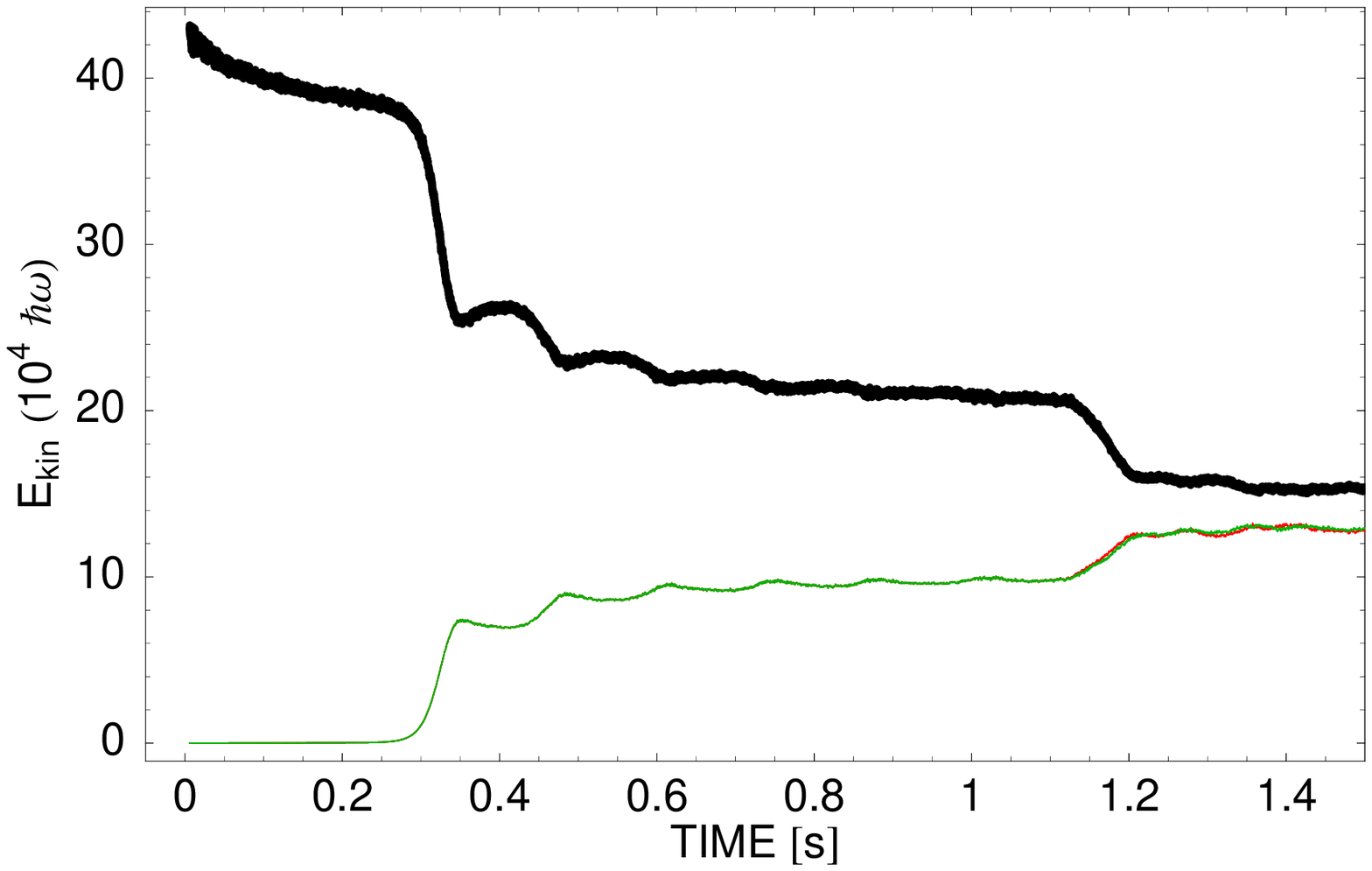}}
\caption[]{(color online) Kinetic energy of $m_F=+1$ (red), $m_F=0$ (black), $m_F=-1$ (green) states as a function of time with dipole interactions (upper panel) and without them (lower panel) for $N=3$x$10^5$ and $\beta=1$. The time of thermalization is shorter if dipolar interactions are included. The external magnetic field is equal to zero.
\label{kin_sf_300k}}
\end{center}
\end{figure}
Within the classical field approximation this amounts to introduce a thermal fraction into the initial state. In all our simulations $20\%$ of atoms are in the thermal clouds, corresponding to $T/T_c \approx 0.58$. 
Moreover, we put a small seed ($0.3\%$ of all atoms) into $m_F=\pm 1$ components what is necessary to initiate a spin dynamics. We study numerically the dynamics leading to a thermal equilibrium. In all simulations we use a grid of 42 points in each direction with spatial step equal to $\delta x =0.6$.

We start our study with a spherically symmetric system, $\beta \equiv \omega_z/\omega_r=1$ (where $\omega_r$ and $\omega_z$ are the radial and axial trap frequencies respectively) of $N=3 \times 10^5$ atoms, $\omega_r=2 \pi \times 100$ Hz and $\mathbf{B}=0$. Due to interactions the initially empty magnetic components become populated and finally all three thermal clouds oscillate  around the same value what signifies a thermal equilibrium, Fig. \ref{nor_sf_300k}. At the equilibrium, populations of thermal fractions of $m_F=\pm1$ components fluctuate independently. Moreover the kinetic energies accumulated in every component saturate and equalize, Fig. \ref{kin_sf_300k}. The total populations of $m_F=\pm 1$ components need not be identical in the presence of dipolar interactions due to a noise triggered spontaneous chiral symmetry breaking.
To our surprise the dipole-dipole interactions, seemingly about order of magnitude smaller than contact spin mixing term, significantly decrease the thermalization time: from 1.2s without dipole-dipole term to 0.35s when this long range term is retained.

\begin{figure}[!htb]
\begin{center}
\resizebox{3.1in}{2.0in}{\includegraphics{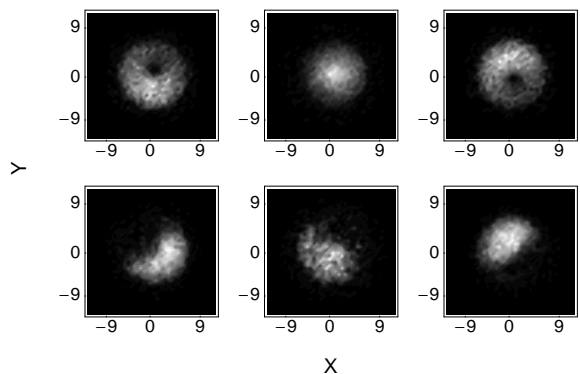}}
\caption[]{The atomic density at $z=0$ plane of $m_F=+1$ (left frame), $m_F=0$ (middle frame) and $m_F=-1$ (right frame) components for the spherically symmetric trap and $N=3 \times 10^5$ atoms at time $t=0.25$s (top) and $t=0.95$s (bottom). The external magnetic field is equal to zero.
\label{chmurki025_sf_300k_42}}
\end{center}
\end{figure}

In order to get a better insight into the origin of the observed enhanced role of the dipole-dipole interactions we examine in more detail a spatial structure of different magnetic components. Typical density profiles are shown in Fig. \ref{chmurki025_sf_300k_42}. At $t=250$ms (upper panel) densities and phases (not shown) of multicomponent spinor wave function indicate the existence of a coreless vortex with winding numbers equal $(-1,0,+1)$ for $m_F=(+1,0,-1)$ respectively. This vortex disappears on the time scale of milliseconds. Let us note that density profiles in $m_F=+1$ and $m_F=-1$ components are not axially symmetric and are rotated with respect to each other by $\pi/2$. The  different magnetic components have small spatial overlap. Similar separation of magnetic phases persists during the evolution, see Fig. \ref{chmurki025_sf_300k_42} (lower panel). 

\begin{figure}[!htb]
\begin{center}
\resizebox{3.4in}{2.0in} {\includegraphics{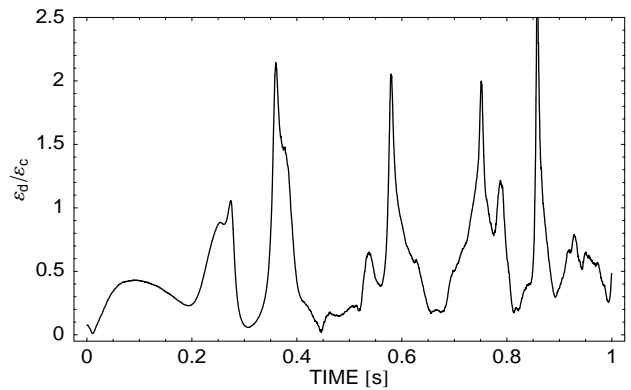}}
\caption[]{Ratio of averaged dipole-dipole energy to the mean value of the contact spin mixing term $\varepsilon_{d}/\varepsilon_{c}$ as a function of time. The external magnetic field is equal to zero.
\label{aspDC}}
\end{center}
\end{figure}
\begin{figure}[!htb]
\begin{center}
\resizebox{3.4in}{2.0in}{\includegraphics{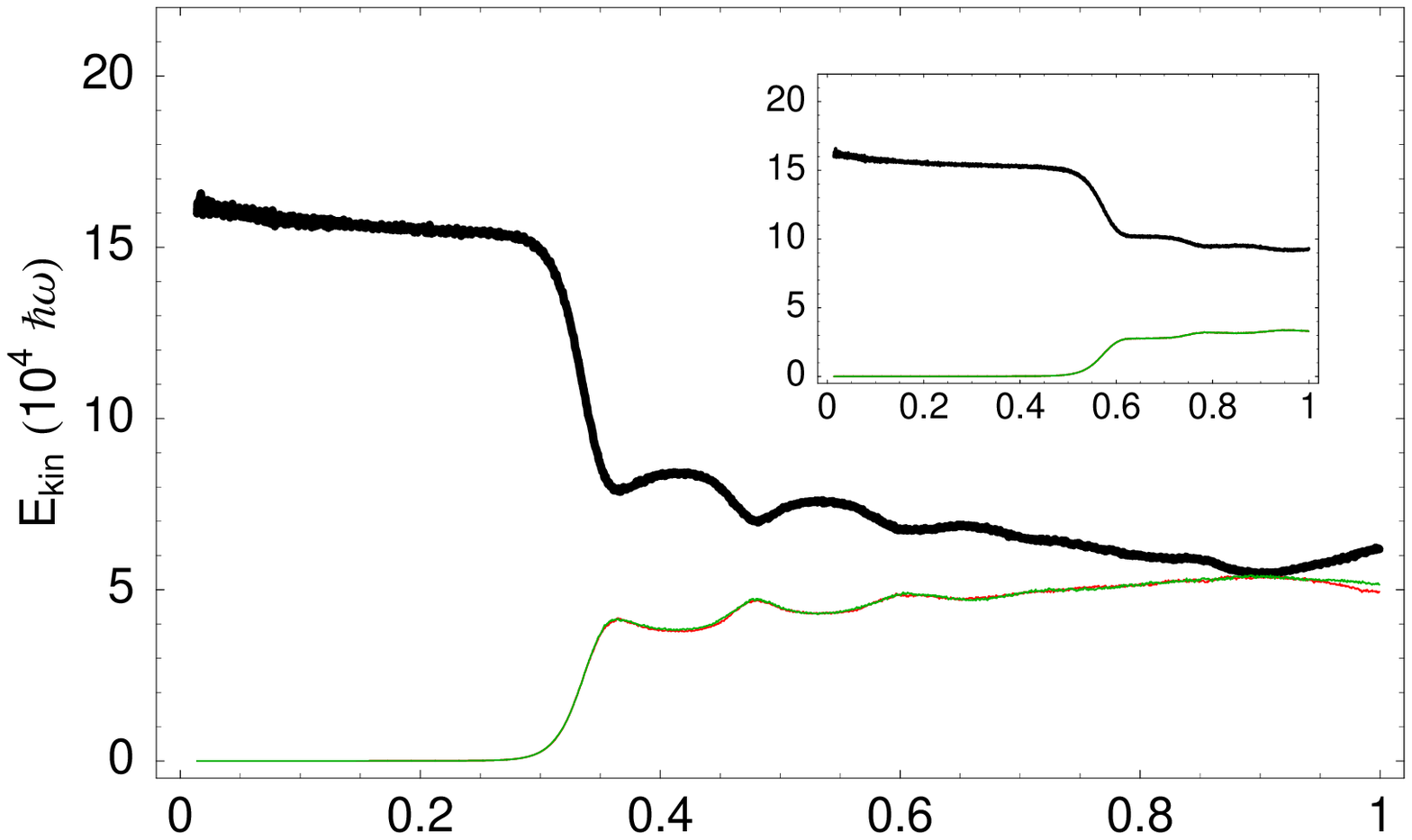}}
\resizebox{3.4in}{2.0in}{\includegraphics{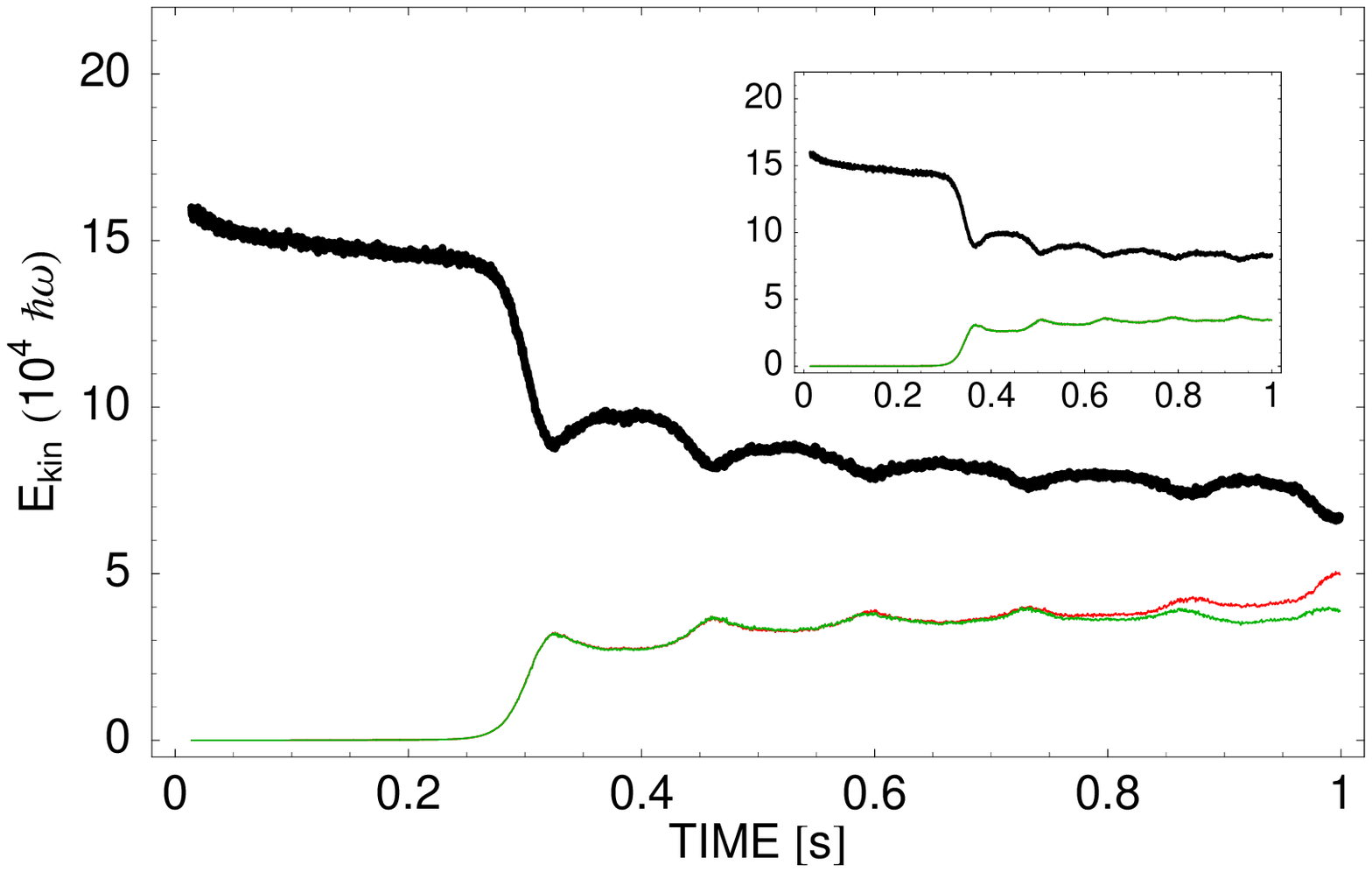}}
\caption[]{(color online) Kinetic energy of $m_F=+1,0,-1$ components for $N=10^5$ atoms as a function of time. Top figure -- the oblate geometry, $\beta=2$, while the lower one -- the prolate shape, $\beta=0.5$. Results were obtained from dynamics generated by the full Hamiltonian - main frame, and by the Hamiltonian without the dipole-dipole term - insets.  The external magnetic field is equal to zero.
\label{kin_nal_100k}}
\end{center}
\end{figure}
The transfer of atoms to initially empty magnetic components arises through dynamical instability.  At initial stages of evolution the contact spin mixing term is responsible for the dynamical instability leading to the formation of domains of opposite magnetization \cite{instability, Hall, Miesner, Kurn}, which only weakly overlap. The dynamically formed spinor wave function cannot be described by a single mode approximation models \cite{Ueda_1,Pu,Kronjager}. The simplest estimation of a magnitude of the contact spin mixing term Eq. (\ref{contact}) and the dipole-dipole energy Eq. (\ref{dipole}) makes use of a mean atomic density. This estimation is not adequate if the magnetic domains are formed because the dipole-dipole interaction is the long range one as opposed to the zero range contact potential. When spin domains are formed the role of the contact term significantly decreases and dipole-dipole interactions become dominant if the spin dynamics is concerned. A mean value of the contact spin-mixing term averaged over the spatial distribution of the spinor wave function is:
\begin{equation}
\varepsilon_{c}=c_2^{} \int d^3r \Psi_1^{*}(\mathbf{r}) \Psi_{-1}^{*}(\mathbf{r}) \Psi_0^{}(\mathbf{r}) \Psi_0^{}(\mathbf{r})\,,
\label{HC10}
\end{equation}
while the dipole-dipole averaged energy reads:
\begin{equation}
\varepsilon_{d}=-\hbar^2 \mu^2 \int d^3r \int d^3r' \Psi_1^{*}(\mathbf{r})  \frac{V(\mathbf{r},\mathbf{r'})}{|\mathbf{r} - \mathbf{r'}|^3}\Psi_0^{}(\mathbf{r})\,,
\label{HD10}
\end{equation}
where $V(\mathbf{r},\mathbf{r'})=3/\sqrt{2} e^{-i\phi} \cos \Theta \sin \Theta (|\Psi_1^{}(\mathbf{r'})|^2-|\Psi_{-1}^{}(\mathbf{r'})|^2)+3/2 e^{-2i\phi} \sin^2 \Theta (\Psi_1^{*}(\mathbf{r'}) \Psi_0^{}(\mathbf{r'})+\Psi_0^{*}(\mathbf{r'}) \Psi_{-1}^{}(\mathbf{r'}))-(1-3/2 \sin^2\Theta) (\Psi_0^{*}(\mathbf{r'}) \Psi_1^{}(\mathbf{r'})+\Psi_{-1}^{*}(\mathbf{r'}) \Psi_{0}^{}(\mathbf{r'}))$, and $\phi$, $\Theta$ are the spherical angles of $\mathbf{R}=\mathbf{r} - \mathbf{r'}$. The ratio of these two terms is shown in Fig. \ref{aspDC}.
\begin{figure}[!htb]
\begin{center}
\resizebox{3.1in}{2.0in}{\includegraphics{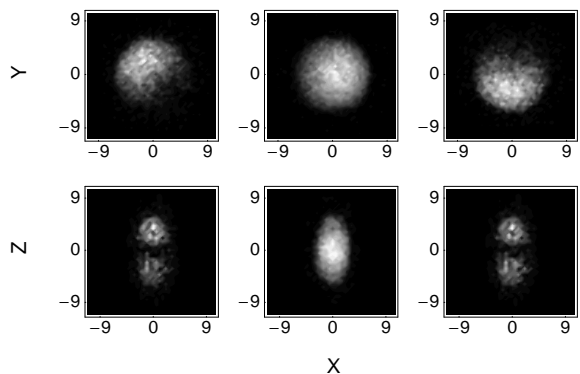}}
\caption[]{Densities of $m_F=+1$ (left frame), $m_F=0$ (middle frame) and $m_F=-1$ (right frame) components for the case of $N=10^5$ atoms. The top frame is for  $\beta=2$ at $t=0.7$s, and the bottom frame is for $\beta=0.5$ at $t=0.35$s. The densities of $m_F=+1$ and $m_F=-1$ components for $\beta=2$ are spatially separated while for $\beta=0.5$ they overlap. 
\label{chmurki095_sf_300k_42}}
\end{center}
\end{figure}
During the evolution the dipole-dipole term is comparable to the contact one and frequently the dipolar interaction becomes dominant. This is the main observation of our study.

The structure of the dynamically unstable modes depends on a geometry of the system. To explore this fact we analyzed two cases,  $\beta=2$ ($\omega_z=2\pi$ x $200$ Hz, $\omega_r=2\pi$ x $100$ Hz) corresponding to an oblate profile and $\beta=0.5$ ($\omega_z=2\pi$ x $100$ Hz and $\omega_r=2\pi$ x $200$ Hz) characteristic for a prolate shape.
In the case of the oblate geometry the dipolar forces significantly accelerate the spin dynamics while no effect is observed for the prolate geometry, Fig. \ref{kin_nal_100k}. The spatial density profiles in Fig. \ref{chmurki095_sf_300k_42} help to understand this difference. For the oblate shape the dynamics leads to the separation of phases and formation of only weakly overlapping domains of opposite magnetization. This is the reason why the role of zero range interaction terms significantly decreases and long range dipole forces start to dominate the spin dynamics. The same dynamics in the case of the prolate geometry favors the formation of almost identical structures of $m_F=1$ and $m_F=-1$ components and spin mixing contact term dominates over the dipole-dipole interaction.

\begin{figure}[!htb]
\begin{center}
\resizebox{3.5in}{2.0in} {\includegraphics{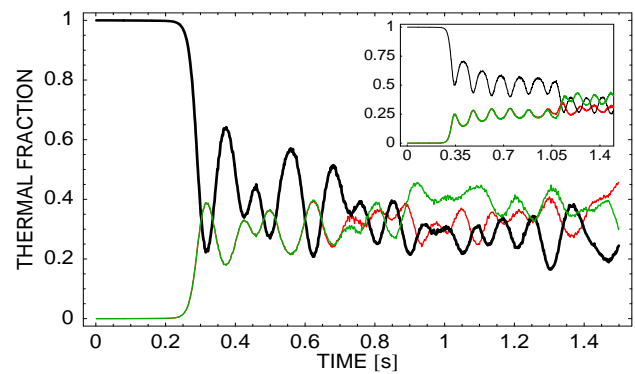}}
\caption[]{(color online) Populations of thermal clouds for $m_F=+1$ (red), $m_F=0$ (black), and $m_F=-1$ (green) states as a function of time with (the main frame) and without (the inset) dipole-dipole interactions for spherical symmetry and for $B_z=100$mG. The time of thermalization at $B_z=100$mG is longer then for $B_z=1$mG, but still remains shorter if dipolar interactions are included. The system thermalizes after $t \approx0.7$s.
\label{THF_100mG}}
\end{center}
\end{figure}

The theoretical studies of dipole-dipole interactions in $F=1$ ${}^{87}$Rb \cite{Ueda_2} show, that it is not easy to observe dipolar effects at non-zero magnetic field. In particular, dipolar effects are easy to observe at magnetic field $\sim 10\mu$G \cite{Gajda_Bongs}. Our study shows, that it is experimentally possible to see dipolar effects by observing time of thermalization of the system. At $T=0K$, the time of thermalization for both type of systems (with and without dipole-dipole interactions) is the same and equals $t\approx0.75$s. For non-zero temperatures ($N_0/N \sim 0.8$, where $N_0$ is a number of condensed atoms and $N$ is the total number of atoms) the time of thermalization is different for both cases. 
\begin{figure}[!htb]
\begin{center}
\resizebox{3.5in}{2.0in} {\includegraphics{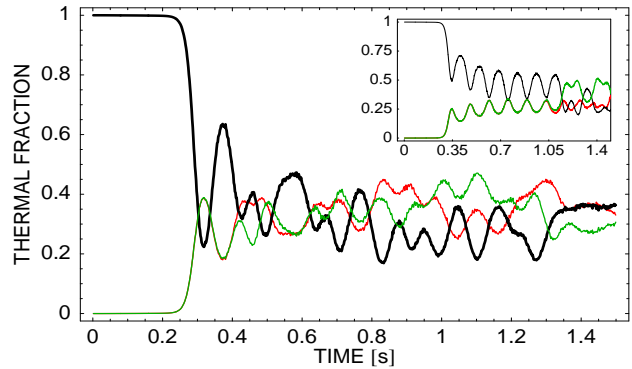}}
\caption[]{(color online) Populations of thermal clouds for $m_F=+1$ (red), $ m_F=0$ (black), and $ m_F=-1$ (green) states as a function of time with (the main frame) and without (the inset) dipole-dipole interactions. The parameters are N=$3$x$10^5$, $\beta=1$, and $B_z=1$mG. The time of thermalization is $t \approx0.4$s.
\label{THF_1mG}}
\end{center}
\end{figure}
\begin{figure}[!htb]
\begin{center}
\resizebox{3.5in}{2.0in} {\includegraphics{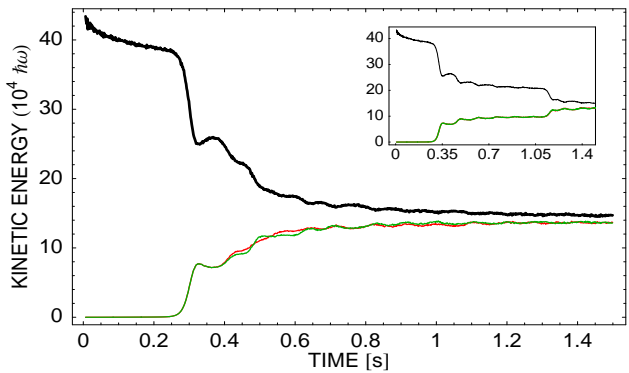}}
\caption[]{(color online)  Kinetic energy of $m_F=+1$ (red), $m_F=0$ (black), $m_F=-1$ (green) states as a function of time with dipole interactions (the main frame) and without them (the inset) for spherical symmetry. The time of thermalization at $B_z=1$mG remains shorter if dipolar interactions are included. 
\label{KIN_1mG}}
\end{center}
\end{figure}
\begin{figure}[!htb]
\begin{center}
\resizebox{3.1in}{2.0in} {\includegraphics{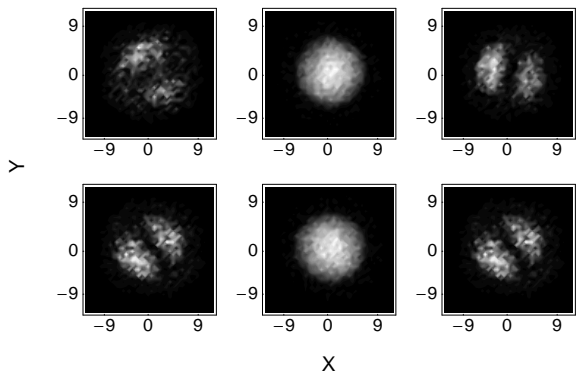}}
\caption[]{(color online) The atomic density at $z=0$ plane of $m_F=+1$ (left frame), $m_F=0$ (middle frame) and $m_F=-1$ (right frame) components for the spherically symmetric trap and $N=3 \times 10^5$ atoms at $B_z=1$mG at time $t=0.45$s with (top) and without (bottom) dipole-dipole interactions.
\label{CLOUDS_1mG}}
\end{center}
\end{figure}
\begin{figure}[!htb]
\begin{center}
\resizebox{3.0in}{3.0in} {\includegraphics{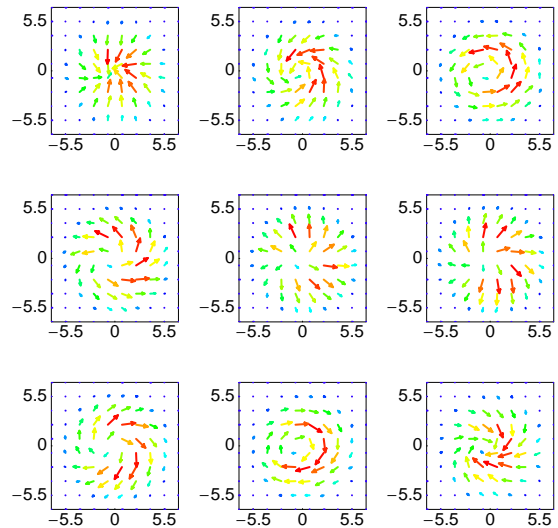}}
\caption[]{(color online) The sequence of frames representing changing of spin textures with dipole-dipole interactions. The value of external magnetic field is $B_z=1$mG and Larmor precession frequency is $\omega_L^{1mG}=6.9$ and $t_L=1.45 \times 10^{-3}$s. The sequence of frames starts from 1.2s and ends after the single cycle.
\label{MOVIE_1mG}}
\end{center}
\end{figure}
The time of thermal equilibrium at $B_z=1$mG (Fig. \ref{THF_1mG}) for spherical trap is almost identical as for $B_z=0$ (Fig. \ref{nor_sf_300k}). The time of thermalization is equal $t\approx0.4$s and is still three times smaller with the dipole-dipole interactions included. 
For $B_z=100$mG, the system is thermalized after $t\approx0.7$s (Fig. \ref{THF_100mG}). 
\begin{figure}[!htb]
\begin{center}
\resizebox{3.5in}{2.0in} {\includegraphics{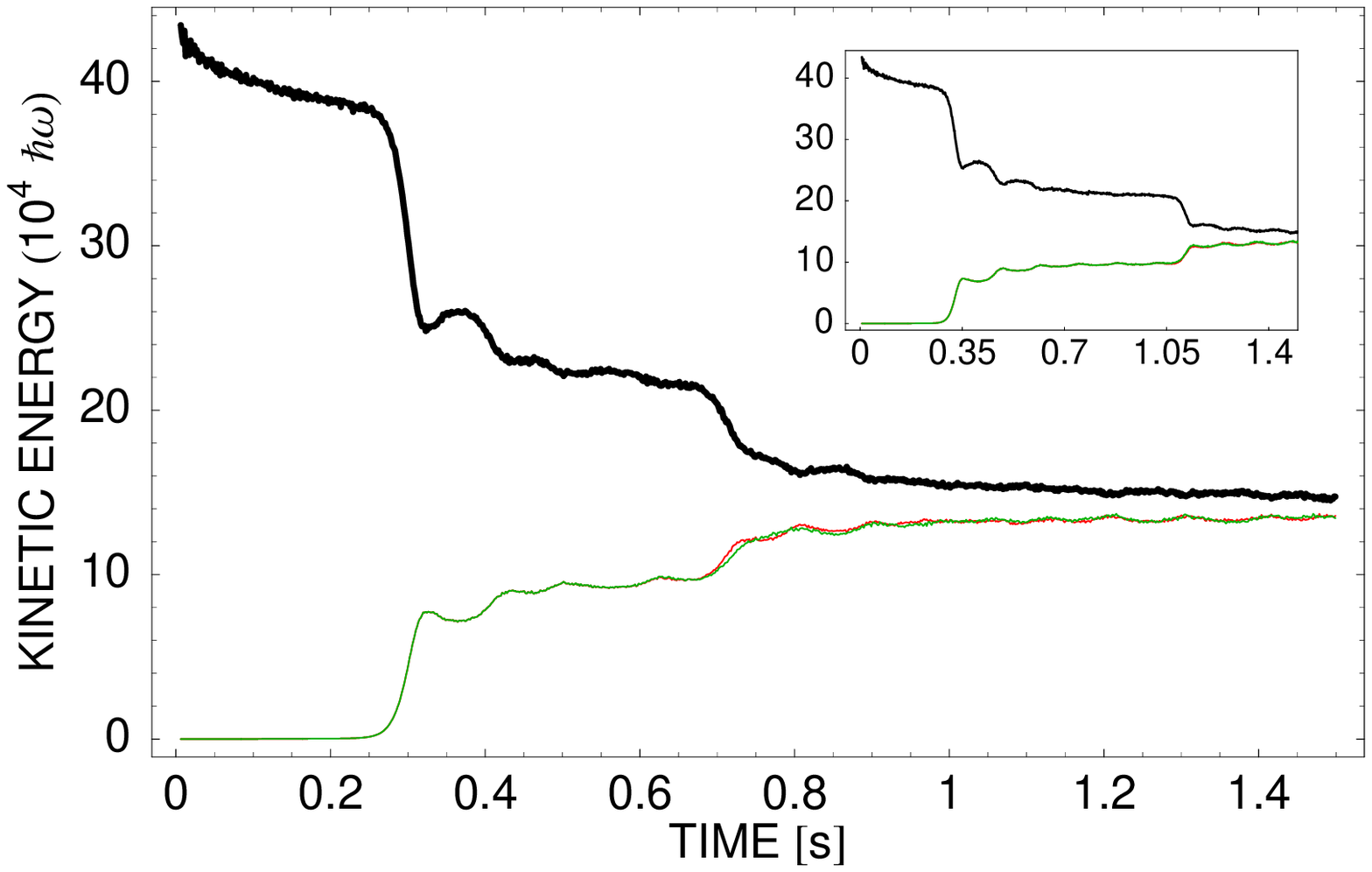}}
\caption[]{(color online)  Kinetic energy of $m_F=+1$ (red), $m_F=0$ (black), $m_F=-1$ (green) states as a function of time with dipole interactions (the main frame) and without them (the inset) for spherical symmetry for magnetic field $B_z=100$mG. The system thermalizes after $t\approx0.7$s.
\label{KIN_100mG}}
\end{center}
\end{figure}
For the without dipolar interactions the system is thermalized after $t\approx1.2$s (this time is the same as without external magnetic field). So, even at higher magnetic fields, we can see dipolar effects by observing the time of thermalization of the system. 
The spatial density profiles (Fig. \ref{CLOUDS_1mG}) show, that in this relatively strong magnetic field the unstable Bogoliubov modes have $m=\pm1$ azimuthal quantum number in a close analogy with  the 2D simulations of \cite{Gawryluk_1}. The latter had no dipole-dipole forces, thus we conclude that the magnetic field tends to mask the dipolar effects \cite{Ueda_2}.

\begin{figure}[!htb]
\begin{center}
\resizebox{3.0in}{3.0in} {\includegraphics{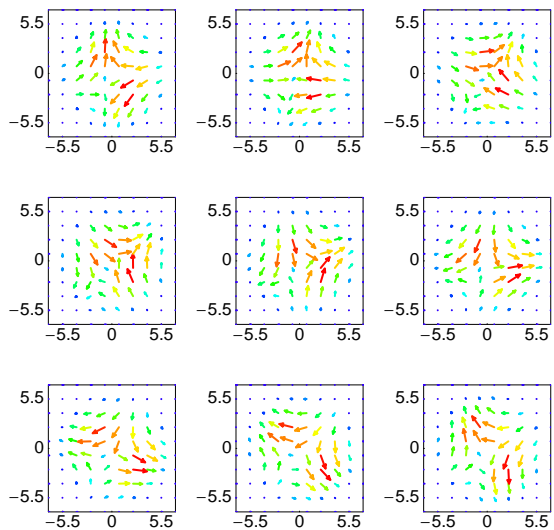}}
\caption[]{(color online) The sequence of frames representing changing of spin textures. The value of external magnetic field is $B_z=100$mG and Larmor precession is $\omega_L^{100mG}=690$ and $t_L=1.45 \times 10^{-5}$s. The sequence of frames starts from 0.75s and ends after the single cycle. The dipole-dipole interactions are included.
\label{MOVIE_100mG}}
\end{center}
\end{figure}

The dynamics of both systems allows to observe rich structures of spin textures. Figure \ref{MOVIE_1mG} shows typical structures of $[F_x,F_y]$ quantity for dipolar case at $B_z=1$mG (Fig. \ref{MOVIE_1mG}) and $B_z=100$mG (Fig. \ref{MOVIE_100mG}).  

For non-dipolar case we can observe similar structures in 'xy' plane. Additionally, rotation of spin domains is observed. During a single rotation, it is possible to observe different structures of spin textures. Using Larmor formula
\begin{equation}
\omega_L=\gamma B_z\,,
\label{e_spin}
\end{equation}
where $\omega_L$ is the Larmor frequency we can calculate the number of rotations per unit time. For example, $\omega_L^{1mG}=6.9$ and $t_L^{1mG}=1.45 \times 10^{-3}$s. For $B_z=100$mG, $\omega_L^{100mG}=690$ and $t_L^{100mG}=1.45 \times 10^{-5}$s which is confirmed by our simulations.
For low magnetic field we observe a different time of spin precession. For example, at $B_z=10\mu$G the rotation period for dipole-dipole interaction case is about two times smaller. It can be explained by replacement of external magnetic field by effective magnetic field which includes external magnetic field, the mean field originating from the spin-exchange and the mean field  originating from the dipole-dipole interactions \cite{Zhang}.

In summary we have shown, that the dipole-dipole interactions play a major role in the dynamics of the ${}^{87}$Rb condensate in $F=1$ hyperfine state: a) the spin changing s-wave scattering length is small (noted before), b) the contact interactions trigger the formation of weakly overlapping magnetic domains, c) because of a) and b) the importance of contact interactions with respect to the long range forces is reduced, d) as a consequence the thermalization time is significantly shorter with than without the dipolar term. We hope, that this findings may be verified experimentally by monitoring dynamics of the thermal clouds of all magnetic components turning off and on the dipole-dipole interactions \cite{Kurn_prl}.

{\bf Acknowledgment}
We thank \L . A. Turski for a stimulating discussion. T.\'S. acknowledges support of the Polish Government Research Founds for 2008. M.B., M.G. and K.R. were supported by Polish Government Research Founds for 2006-2009.

\end{document}